\documentclass[fleqn,usenatbib,onecolumn]{mnras}

\usepackage{newtxtext,newtxmath}
\usepackage{amssymb,amsmath,graphicx}
\usepackage[T1]{fontenc}
\usepackage{ae,aecompl}

\newcommand{\p}{\upartial}

\newcommand{\pc}{\,{\rm pc}}
\newcommand\frakH{\mathfrak{H}}

\newcommand{\rd}{\mathrm{d}} %% derivative
\newcommand{\re}{\mathrm{e}} %% exponentials
\newcommand{\ri}{\mathrm{i}} %% imaginary 

\newcommand{\bfA}{\mathbfit{A}}
\newcommand{\bfB}{\mathbfit{B}}
\newcommand{\bfe}{\mathbfit{e}}
\newcommand{\bfr}{\mathbfit{r}}
\newcommand{\bfv}{\mathbfit{v}}
\newcommand{\bfz}{\mathbfit{z}}

\newcommand\bfmu{\mbox{\boldmath$\mu$}}
\newcommand\bfnu{\mbox{\boldmath$\nu$}}

\title[Phase transition in black-hole star clusters. II.]{Order-disorder phase transition in black-hole star clusters. II. A scale-free cluster}

\author[S. Tremaine]{Scott Tremaine$^{1}$\thanks{E-mail: tremaine@ias.edu}
\\
$^{1}$Institute for Advanced Study, Princeton, NJ 08540, USA\\
}

\pubyear{2015}

\begin{document}
\label{firstpage}
\pagerange{\pageref{firstpage}--\pageref{lastpage}}
\maketitle

\begin{abstract}

The supermassive black holes found at the centres of galaxies are often surrounded by dense star clusters. The ages of these clusters are generally longer than the resonant-relaxation time and shorter than the two-body relaxation time over a wide range of radii. We explore the thermodynamic equilibria of such clusters using a simple self-similar model. We find that the cluster exhibits a phase transition between a high-temperature spherical equilibrium and a low-temperature equilibrium in which the stars are on high-eccentricity orbits with nearly the same orientation. In the absence of relativistic precession, the spherical equilibrium is metastable below the critical temperature and the phase transition is first-order. When relativistic effects are important, the spherical equilibrium is linearly unstable below the critical temperature and the phase transition is continuous. A similar phase transition has recently been found in a model cluster composed of stars with a single semimajor axis. The presence of the same phenomenon in two quite different cluster models suggests that lopsided equilibria may form naturally in a wide variety of black-hole star clusters.

\end{abstract}


\section{Introduction}

Many galaxies contain black holes of mass $10^6$--$10^{10}M_{\sun}$ \citep{kh13} and/or dense nuclear star clusters \citep{gb14} at their centres. For example, the Milky Way contains a black hole of $4.15\times 10^6M_{\sun}$ surrounded by a nuclear star cluster with an average stellar density $\sim 0.4\times 10^5M_{\sun}\pc^{-3}$ in the central parsec \citep{ge10}.

We shall use the term `black-hole star cluster' to denote a stellar system of mass $M_\star$ orbiting a black hole of mass $M_\bullet\gg M_\star$. The structure and dynamics of such clusters govern phenomena such as black-hole formation, tidal disruption events, the fueling history of active galactic nuclei, the gravitational-radiation background from mergers of supermassive black holes, etc. For reviews of this subject see \cite{mer13} and \cite{alex17}.

Direct N-body simulations of black-hole star clusters are challenging  because of the large number of stars (typically $10^5$--$10^7$), the wide range of orbital periods (from hours at the innermost stable circular orbit to $\sim10^5\mbox{\,yr}$ at the half-mass radius), and the requirement to follow the evolution for up to $10^{10}\mbox{\,yr}$ or $10^5$--$10^{13}$ orbits. N-body simulations of the Milky Way star cluster are advancing rapidly in accuracy and realism \citep{bau18,ber18} but still contain far fewer stars than the actual cluster.  Therefore analytic arguments and simulations based on idealized dynamics are essential for studying these systems. 

All star clusters relax through close gravitational encounters between stars, a process called two-body relaxation. The characteristic relaxation time for this process is $t_\mathrm{relax}\sim \Omega^{-1} M_\bullet^2/(M_\star m)$ where $2\pi/\Omega$ is the orbital period and $m$ is a typical stellar mass. A watershed in our understanding of the dynamics of black-hole star clusters was the recognition that over time-scales $\gg t_\mathrm{relax}$ the density distribution of stars near the black hole is determined by the requirement that two-body relaxation generate a constant radial current of orbital energy \citep{pjep72,bw76,fr76,ls77,ck78}. This requirement leads to the Bahcall--Wolf density distribution $n(r)\propto r^{-7/4}$ for a cluster composed of stars of a single mass.

Resonant relaxation, which arises from the orbit-averaged torques between stars in a black-hole cluster, occurs on a time-scale $t_\mathrm{rr}\sim \Omega^{-1} M_\bullet/m$. Resonant relaxation is distinct from two-body relaxation: (i) it leads to diffusion in the eccentricities and angular-momentum vectors of stars but does not affect their semimajor axes \citep{rt96,boa16,st16,bof18}; (ii) if the gravitational potential is dominated by a central black hole and relativistic effects are small, resonant relaxation is much faster than two-body relaxation, by a factor of order $M_\bullet/M_\star$. In the Milky Way's black-hole star cluster, resonant relaxation is faster than two-body relaxation over the range 0.001--$0.1\pc$, by as much as a factor of ten (see Fig.\ 1 of \citealt{kt11} or Fig.\ 4 of \citealt{bof18}). 

The goal of this paper is to investigate the equilibrium structure of black-hole star clusters on time-scales long compared to the resonant-relaxation time but short compared to the two-body relaxation time. On these time-scales the semimajor-axis distribution in the cluster is frozen, but the eccentricity or angular-momentum distribution and the distribution of orbit orientations are in statistical equilibrium. For simplicity we shall make a number of assumptions and simplifications, of which the most important are the following: (i) we consider a self-similar cluster in which the relevant properties of the equilibrium state are independent of semimajor axis; (ii) we assume that all stars at a given semimajor axis have the same mass; (iii) we ignore the consumption or tidal disruption of stars by the central black hole.

With these assumptions, the stars at each semimajor axis can be regarded -- in  sense that we shall make clear -- as a canonical ensemble that is in thermal equilibrium with a heat bath provided by the rest of the cluster. 

We shall show that this system exhibits both first-order and continuous phase transitions between a disordered high-temperature state and an ordered low-temperature state. The disordered state is spherically symmetric, while in the ordered state the stellar orbits have high eccentricity and nearly aligned apsides. In the absence of relativistic precession, the disordered state is usually metastable, so the expected transition may be delayed or absent if the system is sheltered in the equilibrium disordered state.

Over time-scales longer than the two-body relaxation time, the semimajor axis distribution evolves towards the Bahcall--Wolf distribution. This paper does not address the properties of the phase transition in the Bahcall--Wolf cusp. However, \citet[][hereafter TTK19]{ttk19} found a similar phase transition in a black-hole star cluster in which all the stars had the same semimajor axis. The models in these two papers span the range of possible semimajor-axis distributions -- power law versus delta function. Moreover TTK19 reported on Markov Chain Monte Carlo simulations of a Bahcall--Wolf cluster that exhibited a transition to a lopsided state.  These results suggest that the phase transition is a robust property of a wide range of density profiles including the Bahcall--Wolf distribution.

Section \ref{sec:selfsim} describes the self-similar stellar system and how we compute its free energy. The properties of the disordered and ordered equilibria are derived in \S\ref{sec:sph} and \S\ref{sec:non-sph} respectively. The effects of relativistic precession are analyzed in \S\ref{sec:gr}. Section \ref{sec:disc} provides a discussion to set these results in context. 

\subsection{Phase-space variables}

\label{sec:defs}

\noindent
The usual Keplerian orbital elements include the semimajor axis $a$, eccentricity $e$, inclination $I$, argument of periapsis $\omega$, and angle of the ascending node $\Omega$. The position of a particle in its orbit can be specified by the mean anomaly $\ell$, the eccentric anomaly $u$, or the true anomaly $f$. The Delaunay orbital elements  consist of three actions $\Lambda\equiv (GM_\bullet a)^{1/2}$, $L\equiv\Lambda(1-e^2)^{1/2}$, $L_z\equiv L\cos I$ and their conjugate angles $\ell$, $\omega$, and $\Omega$. The actions $L$ and $L_z$ are the angular momentum per unit mass and its $z$-component.

The canonical volume element in phase space is
\begin{align}
\rd\bfmu &= \rd\Lambda \rd L \rd L_z \rd\ell \rd\omega \rd\Omega = \textstyle{\frac{1}{4}}(GM_\bullet)^{3/2} a^{1/2} \rd a \rd e^2\sin I \rd I \rd\ell \rd\omega \rd\Omega\nonumber \\
&=\textstyle{\frac{1}{4}}(GM_\bullet)^{3/2} a^{1/2} \rd a \rd\ell \rd\bfnu
\label{eq:mudef}
\end{align}
where
\begin{equation}
\rd\bfnu\equiv \rd e^2\sin I  \rd I \rd\omega
\rd\Omega.
\label{eq:nudef}
\end{equation}
The (non-canonical) four-dimensional volume element $\rd\bfnu$ is useful because resonant relaxation conserves the semimajor axis and averages over the mean anomaly, so their role is different from that of the other phase-space variables.

\section{A self-similar star cluster surrounding a massive black hole}

\label{sec:selfsim}

We want to construct a stellar system that has the same thermal equilibrium at all radii. By `thermal equilibrium' we mean the state achieved on time-scales long compared to the resonant-relaxation time-scale but short compared to the two-body relaxation time-scale. Then the semimajor axes of the stars are frozen and the stars in each small interval of semimajor axis can be regarded as a subsystem with a fixed number of stars. All these subsystems share a common temperature, specified by $\beta\equiv (k_BT)^{-1}$ where $k_B$ is Boltzmann's constant\footnote{We assume throughout that the temperature is positive, that is $\beta>0$. Negative-temperature states exist but exhibit less interesting behavior.}.  We require that two quantities that govern the equilibrium state are independent of semimajor axis: the number of stars in a fixed logarithmic interval in semimajor axis and the energy of a star in the mean gravitational field of the other stars. These requirements are satisfied if and only if the distribution of stars in semimajor axis and the mass of a star with given semimajor axis satisfy
\begin{equation}
dN =N_0\frac{\rd a}{a}\quad;\quad m(a)=m_0\left(\frac{a}{a_0}\right)^{1/2}
\label{eq:ss}
\end{equation}
where $N_0$, $m_0$, and $a_0$ are constants. The short-term dynamics of this system is not scale-free since the orbital period scales as $a^{3/2}$ while the apsidal precession period scales as $a$, but this is not a concern as the short-term dynamics plays no direct role in determining the thermal equilibrium. We defer discussing the effects of relativistic precession until \S\ref{sec:gr}. 

With these choices the density of stars $\rho \sim r^{-5/2}$. The density and the potential arising from this density are related by Poisson's equation $\nabla^2\Phi=4\pi G\rho$ so we may write 
\begin{equation}
\rho(\bfr)=\frac{1}{r^{5/2}}\sum_{l=0}^\infty\sum_{m=-l}^l A_{lm}Y_{lm}(\theta,\phi), \quad \Phi(\bfr)= -\frac{4\pi G}{r^{1/2}}\sum_{l=0}^\infty\sum_{m=-l}^l\frac{A_{lm}}{(l+\tfrac{1}{2})^2}Y_{lm}(\theta,\phi),
\label{eq:rhopot}
\end{equation}
where $(r,\theta,\phi)$ are the standard spherical coordinates and $Y_{lm}(\theta,\phi)$ is a spherical harmonic. Since the density and potential are real and $Y_{l-m}(\theta,\phi)=(-1)^mY_{lm}^\ast (\theta,\phi)$, we must have $A_{l-m}=(-1)^m A^*_{lm}$.  

The orbit-averaged potential energy of a star is $m\langle\Phi\rangle$ where  $\langle\cdot\rangle$ represents a time average over the orbit. Then
\begin{equation}
m\langle\Phi\rangle=-4\pi Gm\sum_{lm} \frac{A_{lm}}{(l+\tfrac{1}{2})^2}\langle r^{-1/2}Y_{lm}(\theta,\phi)\rangle =-\frac{4\pi G m_0}{a_0^{1/2}}\sum_{lm}
\frac{A_{lm}}{(l+\tfrac{1}{2})^2}W_{lm}(e,I,\omega,\Omega)
\label{eq:a1}
\end{equation}
where 
\begin{equation}
W_{lm}(e,I,\omega,\Omega)\equiv \langle (a/r)^{1/2}Y_{lm}(\theta,\phi)\rangle =\frac{\langle (1+e\cos
f)^{1/2}Y_{lm}(\theta,\phi)\rangle}{(1-e^2)^{1/2}}.
\label{eq:w1def}
\end{equation}
The properties of this function are described in Appendix \ref{app:wu}. Since $m\langle\Phi\rangle$ is real, we can also write 
\begin{align}
m\langle\Phi\rangle&=-\frac{4\pi G m_0}{a_0^{1/2}}\sum_{lm} \frac{\mbox{Re}\big[A_{lm}W_{lm}(e,I,\omega,\Omega)\big]}{(l+\tfrac{1}{2})^2}.
\label{eq:a111}
\end{align}

The mass in a small element of phase space is
\begin{equation}
\rd m = m(a) F(a,e,I,\omega,\Omega)\, \rd\bfmu
\label{eq:dm}
\end{equation}
where $\rd\bfmu$ is given by equation (\ref{eq:mudef}) and $F(a,e,I,\omega,\Omega)$ is the distribution function, defined as the number of stars per unit volume in phase space. In thermal equilibrium $F\propto \exp(-\beta m\langle\Phi\rangle)$. The normalization is determined by the first of equations (\ref{eq:ss}):
\begin{equation}
F(a,e,I,\omega,\Omega)=\frac{2N_0}{\pi (GM_\bullet)^{3/2}a^{3/2}}\frac{\exp(-\beta
  m\langle\Phi\rangle)}{\int \rd\bfnu
  \exp(-\beta m\langle\Phi\rangle)},
\label{eq:df}
\end{equation}
where $\rd\bfnu$ is defined by equation  (\ref{eq:nudef}). 

The mass distribution given by equations (\ref{eq:dm}) and (\ref{eq:df}) must self-consistently generate the density distribution (\ref{eq:rhopot}). To derive this condition we re-write the first of equations (\ref{eq:rhopot}) as
\begin{align}
A_{lm}&={r'}^{1/2}\int \rho(\bfr)\rd\bfr\,Y_{lm}^*(\theta,\phi)\delta(r-r').
\end{align}
We replace $\rho(\bfr)\rd\bfr$ by $\rd m$ using equations (\ref{eq:mudef}), (\ref{eq:nudef}), (\ref{eq:ss}), (\ref{eq:dm}), and (\ref{eq:df}): 
\begin{align}
A_{lm}&=\frac{N_0m_0{r'}^{1/2}}{2\pi a_0^{1/2} 
\int \rd\bfnu\,\exp(-\beta
        m\langle\Phi\rangle)} \int \frac{\rd a \rd\ell \rd\bfnu}{a^{1/2}}\, 
  Y_{lm}^*(\theta,\phi)\delta[r(a,e,\ell)-r']\exp(-\beta m\langle\Phi\rangle)\nonumber \\
 &=\frac{N_0m_0}{2\pi a_0^{1/2}\int
   \rd\bfnu   \exp(-\beta m\langle\Phi\rangle)} \int \rd\ell \rd\bfnu\,
  \frac{(1+e\cos f)^{1/2}}{(1-e^2)^{1/2}}Y_{lm}^*(\theta,\phi)\exp(-\beta m\langle\Phi\rangle)\nonumber \\
 &=\frac{N_0m_0}{a_0^{1/2} 
   \int \rd\bfnu \,  \exp(-\beta m\langle\Phi\rangle)   }\!\int \!\rd\bfnu\,
  W^*_{lm}(e,I,\omega,\Omega)\exp(-\beta m\langle\Phi\rangle).
\label{eq:a2}
\end{align}
Equations (\ref{eq:a1}) and (\ref{eq:a2}) provide a set of non-linear equations for $\{A_{lm}\}$ that define the thermal equilibrium state at any inverse temperature $\beta$. We call these solutions $\bfA^\mathrm{eq}(\beta)=\{ A_{lm}^\mathrm{eq}(\beta)\}$. 

An alternative and more powerful approach is to regard the multipole moments $\bfA$ as arbitrary order parameters, and to calculate the free energy of the system as a function of these parameters. This calculation is carried out in Appendix \ref{app:free} and yields a free energy per unit mass 
\begin{align}
g(\beta,\bfA)&=-\frac{1}{\beta}\log\int\! \rd\bfnu\,\exp\bigg[-\beta m\langle\Phi\rangle -\frac{2\pi G\beta}{N_0}\sum_{lm}\frac{|A_{lm}|^2}{(l+\tfrac{1}{2})^2}\bigg]\nonumber \\&=-\frac{1}{\beta}\log\int\! \rd\bfnu\,\exp\bigg[ \frac{4\pi G m_0\beta }{a_0^{1/2}}\sum_{lm}\frac{\mbox{Re}\,\big[A_{lm}W_{lm}(e,I,\omega,\Omega)\big]}{(l+\tfrac{1}{2})^2}-\frac{2\pi G\beta}{N_0}\sum_{lm}\frac{|A_{lm}|^2}{(l+\tfrac{1}{2})^2}\bigg];
\label{eq:free1}
\end{align}
the second line is equivalent to the first because of equation (\ref{eq:a111}).  We show in the Appendix (eq.\ \ref{eq:aeqdef}) that the self-consistency condition (\ref{eq:a2}) is the same as the condition for an extremum of the free energy.

Characterizing the thermodynamic equilibria through the free energy offers several advantages: (i) The free energy can be used to determine thermodynamic stability: minima of the free energy are stable, while saddle points or maxima are unstable. (ii) Finding minima of a single function in a multi-dimensional space is numerically easier than finding simultaneous roots of multiple functions. (iii) Derivatives of the free energy with respect to the temperature have a simple thermodynamic interpretation (e.g., eq.\ \ref{eq:thermo}). (iv) Spherically symmetric equilibria have $A^\mathrm{eq}_{lm}=0$ for all $l>0$. To find non-spherical equilibria we must truncate the sum over $l$ and $m$ in equation (\ref{eq:free1})  at some $l_\mathrm{max}$. Because of this truncation, we only determine an upper limit to the free energy, although one that should be close to the true energy if $l_\mathrm{max}$ is large enough (see discussion following eq.\ \ref{eq:free34}). If this upper limit is less than the free energy of the spherical configuration with the same temperature, then the spherical equilibrium \emph{must} be unstable or metastable, no matter what value we choose for $l_\mathrm{max}>0$.

The expression for the free energy can be simplified by choosing units such that
\begin{equation}
a_0=1, \quad N_0m_0=1, \quad 2\pi Gm_0=1.
\label{eq:units}
\end{equation}
Then
\begin{equation}
 g( \beta, \bfA)=-\frac{1}{
  \beta}\log\int \rd e^2\sin I \rd I \rd\omega \rd\Omega\,\exp \bigg(\sum_{lm}
\frac{\beta}{(l+\tfrac{1}{2})^2}\big[2\,\mbox{Re\,}[A_{lm}W_{lm}(e,I,\omega,\Omega)] -|A_{lm}|^2\big]\bigg).
\label{eq:free2}
\end{equation}

The spherical-harmonic expansion (\ref{eq:rhopot}) is not the only possible set of basis functions for the density and potential. Suppose we parametrize the density and potential by functional forms with parameters $\bfB$, that is 
\begin{equation}
\rho(\bfr)=\frac{1}{r^{5/2}}R(\bfB;\theta,\phi), \quad \Phi(\bfr)=-\frac{4\pi G}{r^{1/2}}F(\bfB;\theta,\phi).
\end{equation}
Here the two functions satisfy Poisson's equation
\begin{equation}
\frac{1}{\sin\theta}\frac{\p }{\p\theta}\sin\theta\frac{\p F}{\p\theta}
+\frac{1}{\sin^2\theta}\frac{\p^2 F}{\p\phi^2}-\textstyle{\frac{1}{4}}F+
R=0.
\end{equation}
From equation (\ref{eq:rhopot}) we have
\begin{align}
\frac{1}{\Gamma}\int_\Gamma \rd\bfr\,\Phi(\bfr)\rho(\bfr)&=-4\pi
     G\sum_{lm}\frac{|A_{lm}|^2}{(l+\tfrac{1}{2})^2} =-4\pi G\int\sin\theta
   \rd\theta \rd\phi\,F(\bfB;\theta,\phi)R(\bfB;\theta,\phi).
   \label{eq:phieval}
\end{align}
Here $\int_\Gamma$ denotes an integral over a spherical annulus with inner and outer radii differing by a factor $\exp(\Gamma)$. Using the last two expressions to eliminate $|A_{lm}^2|$ from equation (\ref{eq:free1}), the free energy can be rewritten as
\begin{equation}
 g( \beta, \bfB)=\frac{2\pi G}{N_0}\int
 \sin\theta \rd\theta         \rd\phi\,F(\bfB;\theta,\phi)R(\bfB;\theta,\phi) -\frac{1}{ \beta}\log\int \rd e^2\sin I \rd I \rd\omega \rd\Omega\,\exp \Big[4\pi G\beta m_0a_0^{-1/2}\langle (a/r)^{-1/2}F(\bfB;\theta,\phi)\rangle \Big].
\label{eq:free3}
\end{equation}
The advantage of the spherical-harmonic expansion (\ref{eq:free1}) is that the evaluation of equation (\ref{eq:phieval}) in terms of the order parameters $A_{lm}$ does not require a numerical integration; the advantage of other basis-function expansions is that they may represent the density and potential accurately with a much smaller number of order parameters. 

\begin{figure}
\includegraphics[width=\columnwidth]{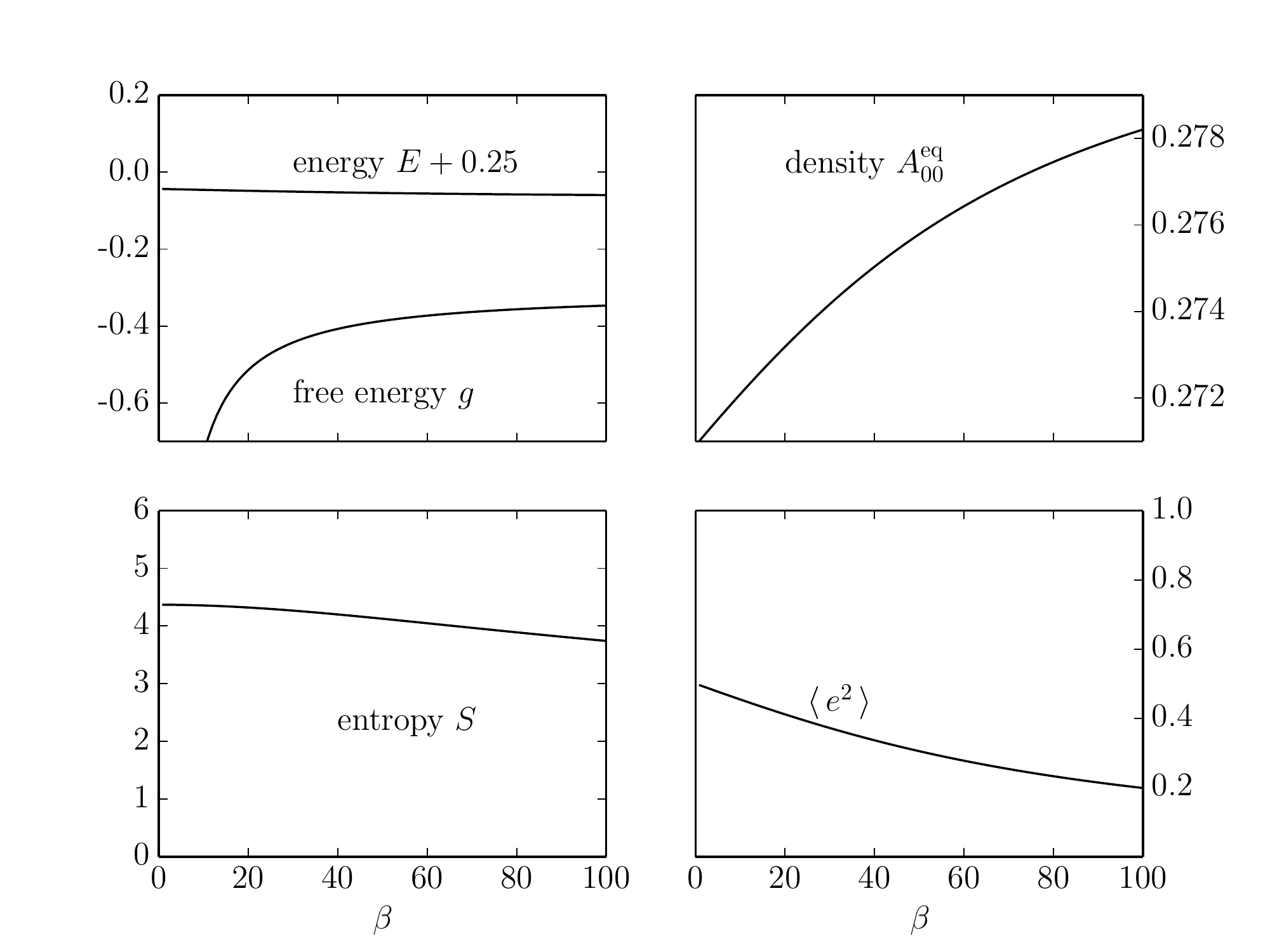}
\caption{Spherically symmetric equilibria of the self-similar system as a function of inverse temperature $\beta$, in the units of equation (\ref{eq:units}). Top left: energy $E(\beta)$ (eq.\ \ref{eq:energydef}) and free energy $g(\beta)=g(\beta,A_{00}^\mathrm{eq})$ (eq.\ \ref{eq:freesph}). A constant has been added to the energy to avoid overlap with the curve for free energy. Top right: density coefficient $A_{00}^\mathrm{eq}$ (eq.\ \ref{eq:rhopot}). Bottom left: entropy $S(\beta)$ (eq.\ \ref{eq:sdef}). Bottom right: mean-square eccentricity $\langle e^2\rangle$. }
\label{fig:sph}
\end{figure} 

\section{Spherical equilibria}

\label{sec:sph}

In a spherically symmetric system $A_{lm}=0$ unles11s $l=m=0$, $A_{00}$ must be real, and $W_{00}(e,I,\omega,\Omega)=U_0(e)/\sqrt{4\pi}$ (eq.\ \ref{eq:wdef}) where $U_0(e)$ is given by equation (\ref{eq:u0}). Equation (\ref{eq:free2}) for the free energy simplifies to 
\begin{equation}
 g( \beta, A_{00})=-\frac{1}{ \beta}\log 8\pi^2\!\!\int
\rd e^2\,\exp 4\beta\big[\pi^{-1/2}U_0(e) A_{00}-A_{00}^2\big]. 
\label{eq:freesph}
\end{equation}

In Fig.\ \ref{fig:sph} we plot the free energy $g(\beta)=g[\beta,A_{00}^\mathrm{eq}(\beta)]$,  the energy $E(\beta)$ (eq.\ \ref{eq:energydef}), the coefficient $A_{00}^\mathrm{eq}$, and the entropy $S$ (eq.~\ref{eq:sdef}) as functions of the inverse temperature $\beta$. These curves satisfy the usual thermodynamic relations (eq.\ \ref{eq:thermo}) such as $g=E-S/\beta$. In the lower right panel we plot the mean-square eccentricity $\langle e^2\rangle$  as a function of $\beta$. At infinite temperature ($\beta=0$) the stars are uniformly distributed in phase space at a given semimajor axis and the mean-square eccentricity is therefore $\langle e^2\rangle=\tfrac{1}{2}$. As the temperature drops, the mean-square eccentricity declines as well, until as $\beta\to\infty$ all of the stars are on circular orbits.

From equation (\ref{eq:rhopot}) the density is $\rho(r)= A_{00}/(\sqrt{4\pi}r^{5/2})$. As seen in the top right panel of Fig.\ \ref{fig:sph}, the function $A_{00}$ is a very slow function of the inverse temperature $\beta$, varying by only 4\% (from 0.27091 to 0.28210) between $\beta=0$ and $\beta\to\infty$ (for comparison the mass of stars per unit semimajor axis is $\rd M/\rd a=4\pi a^2\rho(a)$ where $\rho(a)=1/(4\pi a^{5/2})=0.28210/(\sqrt{4\pi}a^{5/2})$ at all temperatures). 

\begin{figure}
\includegraphics[width=0.9\columnwidth]{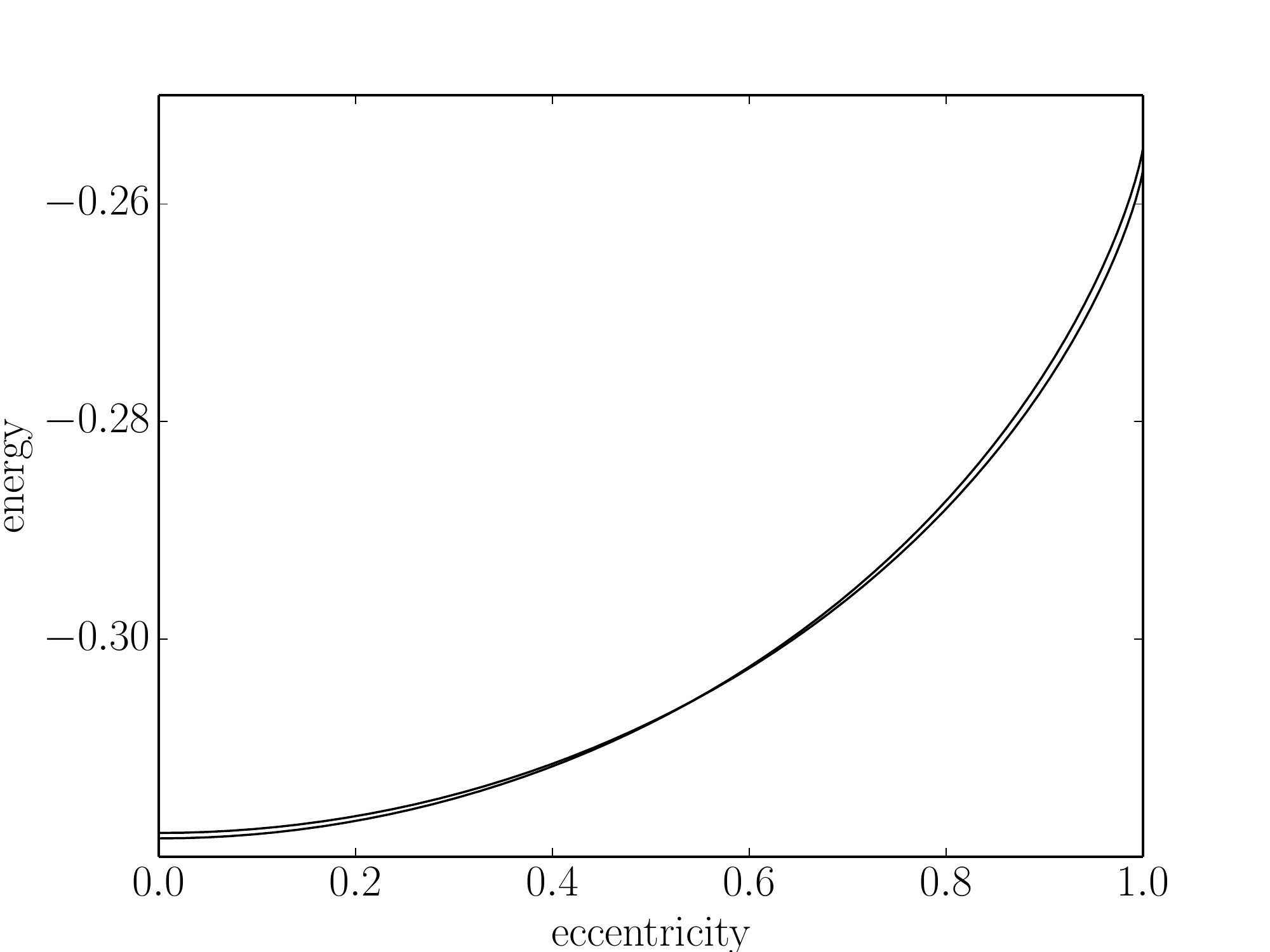}
\caption{The energy of a star in the spherically symmetric self-similar system. The curves are determined using equation (\ref{eq:hdef}) in the units defined by equations (\ref{eq:units}), with $A_{lm}=0$ for $l>0$. The energies are shown for the limiting values $A_{00}=0.27091$ ($\beta\to 0$) and $A_{00}=0.28210$ ($\beta\to\infty$); for intermediate temperatures the energy should lie between these two curves.}
\label{fig:eee}
\end{figure} 

In Fig.\ \ref{fig:eee} we plot the energy $\frakH$ of a star as a function of its eccentricity\footnote{Note that $\frakH=m\langle\Phi\rangle+\mbox{constant}$ for fixed values of $A_{lm}$, a consequence of the lack of a kinetic energy term in the Hamiltonian.}. We have used equation (\ref{eq:hdef}) with $A_{lm}=0$ for $l>0$; we plot the results for the two limiting values of $A_{00}$ at very low and very high temperature, which are nearly the same. The curves show that the energy grows with eccentricity, consistent with our earlier observation that the mean-square eccentricity shrinks as the system is cooled. 

In this paper we do not consider the loss of stars to the central black hole. However, the qualitative effect of this process is straightforward to estimate. The stars that are lost are on near-radial orbits and therefore have the largest energies according to Fig.\ \ref{fig:eee}. Therefore the average energy per unit mass of the surviving stars decreases, which corresponds to decreasing temperature or increasing $\beta$. In other words the self-similar system evolves to the right -- to cooler temperatures -- in the panels of Fig.\ \ref{fig:sph} as stars are destroyed by the black hole, thereby making the phase transition more likely.

\section{Non-spherical equilibria}

\label{sec:non-sph}

\begin{figure}
\includegraphics[width=\columnwidth]{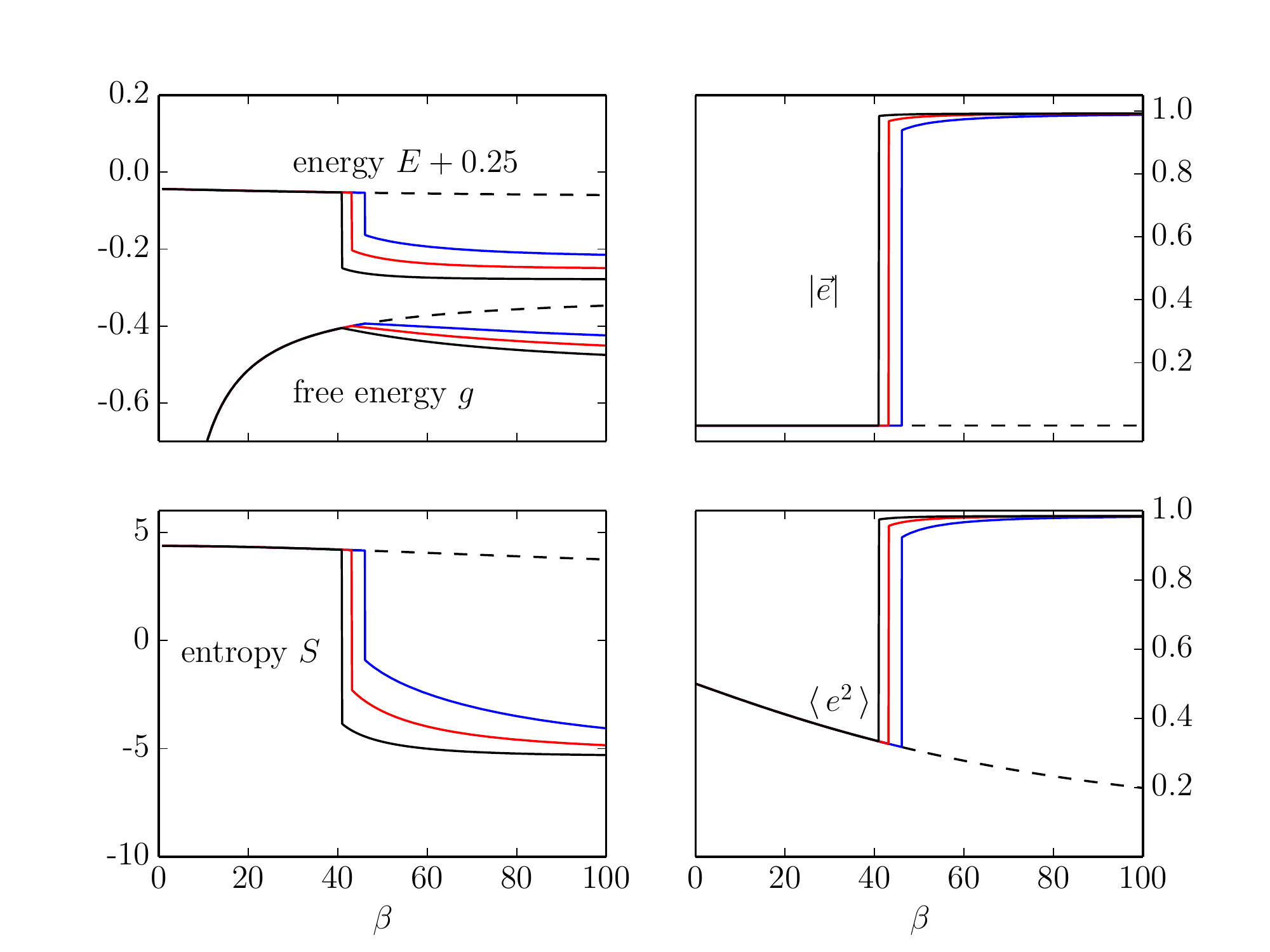}
\caption{Equilibria of the self-similar system as a function of inverse temperature $\beta$, in the units of equation (\ref{eq:units}). Top left: energy $E(\beta)$ (eq.\ \ref{eq:energydef}) and free energy $g(\beta)=g[\beta,\bfA^\mathrm{eq}(\beta)]$ (eq.\ \ref{eq:free1}). A constant has been added to the energy to avoid overlap with the curves for free energy. Top right: magnitude of the mean eccentricity vector (eq.\ \ref{eq:evec}). Bottom left: entropy $S(\beta)$ (eq.\ \ref{eq:sdef}). Bottom right: mean-square eccentricity $\langle e^2\rangle$. Three curves are shown in each panel: the spherical-harmonic expansion with $l_\mathrm{max}=6$ (blue) and 10 (red), and the parametrized model of equations (\ref{eq:free3}) and (\ref{eq:ell}) (black). The dashed curves show the spherical systems of Fig.\ \ref{fig:sph}. As discussed in Appendix \ref{app:free}, the non-spherical calculations yield upper limits to the equilibrium free energy. }
\label{fig:non-sph}
\end{figure} 

\subsection{Numerical methods}

The summation over multipole moments must be truncated at some $l_\mathrm{max}$. Typically we try several values up to $l_\mathrm{max}=10$ to explore the rate of convergence. 

One consequence of using a spherical-harmonic expansion is that the density can be negative at some angular positions. In the most extreme cases the negative densities are no more than a few percent of the maximum density and we do not believe that this level of inaccuracy  compromises our results. 

Because the system is invariant under rotations, we must specify the orientation of any non-spherical equilibrium for the problem to be well-posed. This can be done without loss of generality by requiring 
\begin{equation}
A_{11}=A_{1-1}=0 \quad\mbox{and}\quad A_{10}> 0.
\label{eq:conv}
\end{equation}

Because $A_{l-m}=(-1)^m A^*_{lm}$ we need only track multipoles with $m\ge0$. We have searched for  non-axisymmetric equilibria but have not found any, so we restrict our attention to axisymmetric equilibria by requiring $A_{lm}=0$ for $m\not=0$. We find that $A_{lm}^\mathrm{eq}$ is real in all of our experiments.

The most time-consuming numerical task is the quadrature over $\rd\bfnu$ in equation (\ref{eq:free1}), which has four dimensions, or three if we assume axisymmetry. We carry out this quadrature using the extended midpoint rule on a fixed grid that is uniform in $e^2$, $\cos I$, $\omega$, and $\Omega$, typically containing 32 points in each dimension. The physical reason for not using a more sophisticated quadrature rule is that this method corresponds to an {\em exact} calculation of the free energy on a discrete phase space in which stars are found only at the grid points. This feature makes comparison of the free energy in different configurations more straightforward. 

Since we are interested in lopsided systems, we parametrize the departure from spherical symmetry by the mean eccentricity vector. The eccentricity or Runge--Lenz vector of a Keplerian orbit is defined in terms of the position $\bfr$ and velocity $\bfv$ to be
\begin{equation}
    \bfe\equiv \frac{\bfv\times(\bfr\times\bfv)}{GM_\bullet}-\hat\bfr.
    \label{eq:evec}
\end{equation}
The eccentricity vector points towards periapsis and has magnitude equal to the scalar eccentricity $e$. 

Numerical methods find local minima of the free energy but do not guarantee that these are global minima. It is important to search for local minima using several different starting points, to avoid being trapped at a minimum that is local but not global. 

\subsection{Results}

\label{sec:results}

Fig.\ \ref{fig:non-sph} is the analog of Fig.\ \ref{fig:sph}, without the restriction to spherical symmetry. The dashed curves show the spherically symmetric equilibria from Fig.\ \ref{fig:sph} and the blue and red curves show the equilibria for $l_\mathrm{max}=6$ and 10. In the upper right panel we have replaced  the density coefficient $A_{00}$ with the magnitude of the mean eccentricity vector
\begin{equation}
\label{eq:vece}
|\langle \bfe\rangle|=|\langle\bfe\rangle \cdot\hat\bfz|=\frac{\int \rd e^2\sin I dI \rd\omega \, F(a,e,I,\omega,\Omega)\, e\sin I\sin\omega }{\int \rd e^2\sin I dI \rd\omega \, F(a,e,I,\omega,\Omega)},
\end{equation}
which can range from 0 in spherically symmetric systems to unity when all of the stars are on radial orbits with a common line of apsides.

As the system cools ($\beta$ increases) it undergoes a phase transition from a spherically symmetric or disordered state to an ordered state, in which all of the orbits have eccentricity near unity and their apsidal lines are nearly aligned, i.e., $\langle e^2\rangle\simeq 1$ and $|\langle \bfe\rangle|\simeq 1$. The exact location of the phase transition is difficult to determine accurately because the spherical-harmonic expansion converges slowly in the ordered state.

We have also computed the minimum of the free energy (\ref{eq:free3}) for functions of the form
\begin{equation}
F(\bfB;\theta)=B_0 - B_1K(\theta_s)\ \mbox{where}\ 
\theta_s=\sqrt{\theta^2+B_3^2}+\pi-\sqrt{\pi^2+B_3^2}.
\label{eq:ell}
\end{equation}
Here $K$ is a complete elliptic integral. The motivation for this form is that $-K(\theta)/r^{1/2}$ is the gravitational potential from a wire oriented along the positive $z$-axis with linear density $\propto r^{-1/2}$ and therefore should approximate the potential from the ordered state. The variable $\theta_s$ is a softened version of $\theta$ that avoids the logarithmic singularity at $\theta=0$; in the typical case where $0<B_3\ll\pi$ then $\theta_s$ varies from $B_3(1-\tfrac{1}{2} B_3/\pi)$ to $\pi$ as $\theta$ varies from $0$ to $\pi$.  The results are shown in Fig.\ \ref{fig:non-sph} for $B_3=0.001$ and confirm that there is a phase transition near $\beta=40$, $g=-0.41$, $\langle e^2\rangle^{1/2}=0.58$. More general functions $F(\bfB,\theta)$ would allow a more accurate location of the phase transition.

The phase transition is first-order, that is, there is a discontinuity in the energy (a latent heat). The two phases can co-exist, although this phenomenon is not captured by the mean-field approximation  that we are using here. Thus a black-hole star cluster could contain isolated groups of stars on high-eccentricity orbits with aligned apsides even though most of the stars are spherically distributed. 

\begin{figure}
\includegraphics[width=0.9\columnwidth]{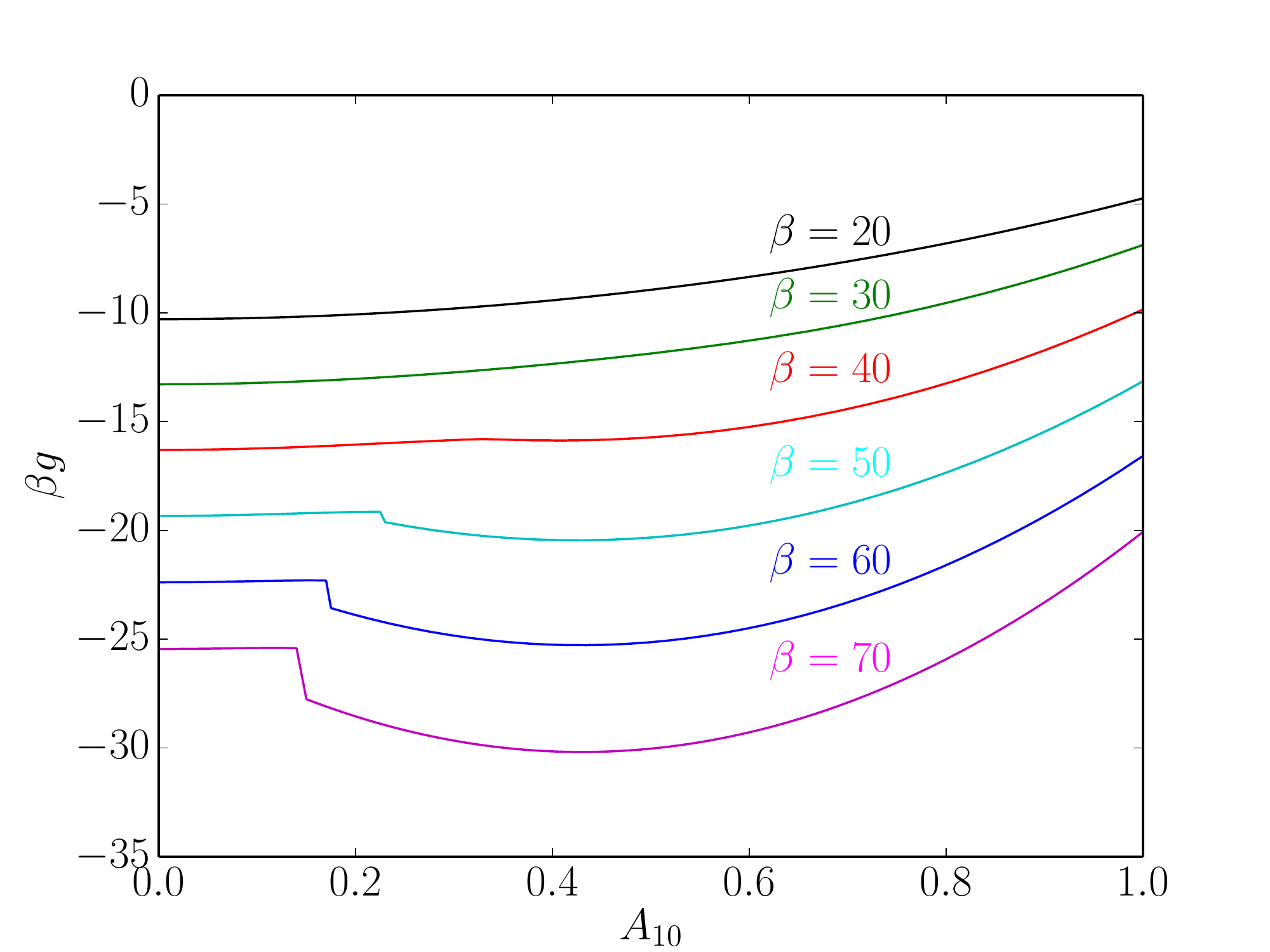}
\caption{Minimum of the free energy as a function of the magnitude of the order parameter $A_{10}$. }
\label{fig:constrain}
\end{figure}

The nature of the transition is illuminated by Fig.\ \ref{fig:constrain}. To construct this plot we have minimized the free energy $g(\beta,\bfA)$ at several inverse temperatures $\beta$ subject to the constraint that the $l=1$ density multipole $A_{10}$ has the value given on the horizontal axis. We see that for $\beta\loa 40$ there is a single minimum that occurs for the spherical state ($A_{10}=0$), while for larger inverse temperatures the state with minimum free energy is non-spherical. Note that the spherical equilibrium is metastable at all temperatures plotted, that is, it is a local minimum of the free energy as a function of the order parameter $A_{10}$. 

\section{Relativistic precession}

\label{sec:gr}

\begin{figure}
\includegraphics[width=\columnwidth]{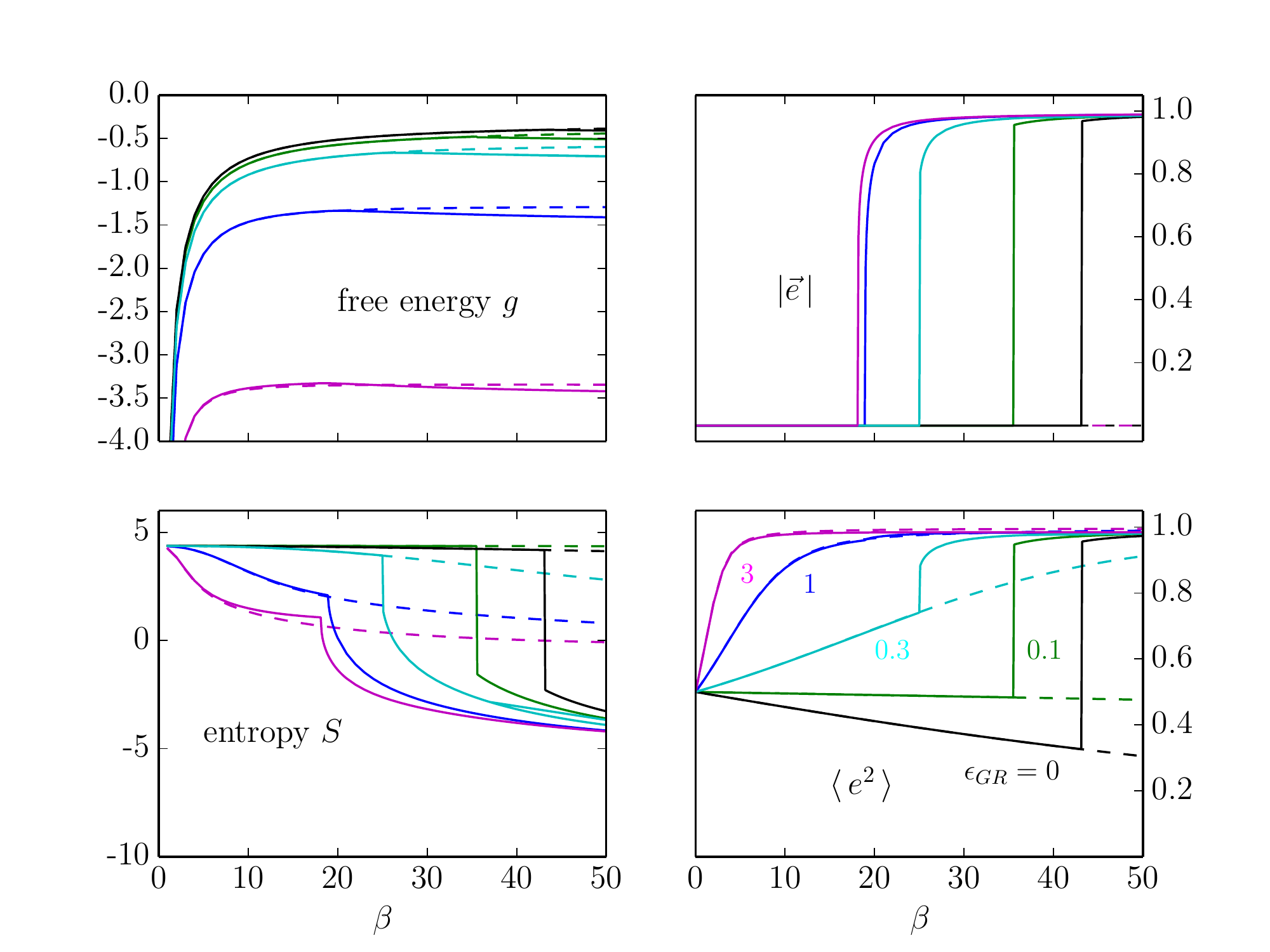}
\caption{Similar to Fig.\ \ref{fig:non-sph}, except that the Hamiltonian in the free energy has an additional term $\widetilde{H}_\mathrm{gr}$ (eq.\ \ref{eq:gr0}) that approximately represents the effects of relativistic precession. In these plots $l_\mathrm{max}=10$, and the softening parameter in the relativistic Hamiltonian is $e_\mathrm{gr}=1.1$. The strength of the relativistic precession relative to precession due to self-gravity is parametrized by $\epsilon_\mathrm{gr}$ (eq.\ \ref{eq:epsgr}), with $\epsilon_\mathrm{gr}=0$ (black), 0.1 (green), 0.3 (cyan), 1 (blue) and 3 (magenta). The dashed lines show the analogous curves for spherical equilibria ($l_\mathrm{max}=0$).}
\label{fig:gr}
\end{figure} 

The calculations so far ignore apsidal precession due to all effects other than the self-gravity of the stars. The most important of these is relativistic precession; for example, in the Milky Way relativistic precession dominates inside about $0.005\pc$ from the black hole \citep{kt11}. 

The apsidal precession rate due to general relativity is
\begin{equation}
\dot\omega=\frac{3(GM_\bullet)^{3/2}}{c^2a^{5/2}(1-e^2)},
\label{eq:grprec}
\end{equation}
where $c$ is the speed of light. Since $\dot\omega=\p H/\p L$, this precession can be produced by a Hamiltonian of the form\footnote{This Hamiltonian produces the wrong relativistic correction for the mean motion, $\dot\ell=-\p H_\mathrm{gr}/\p \Lambda$, but this shortcoming has no effect on our results.}
\begin{equation}
H_\mathrm{gr}=-\frac{3(GM_\bullet)^2}{c^2a^2(1-e^2)^{1/2}}=-\frac{3(GM_\bullet)^4}{c^2\Lambda^3L}.
\label{eq:hgr}
\end{equation}
Interpreting the effect of the relativistic Hamiltonian on our model system is not straightforward, for two reasons. First, the Hamiltonian (\ref{eq:hgr}) breaks the scale invariance. Second, if the temperature is positive, integrals of the form $\int \rd e^2\exp(-\beta H_\mathrm{gr})$ that appear in the free energy or partition function diverge. This divergence is unphysical because (i) orbits with high eccentricity precess rapidly, so resonant relaxation is ineffective; (ii) stars on radial orbits are consumed by the central black hole. 

For a first look at the effects of relativistic precession we evade these problems by replacing the Hamiltonian (\ref{eq:hgr}) by 
\begin{equation}
\label{eq:gr0}
\widetilde{H}_\mathrm{gr}=-\frac{3(GM_\bullet)^2}{c^2a_0^2(e_\mathrm{gr}^2-e^2)^{1/2}}=-\frac{3\epsilon_\mathrm{gr}GN_0m_0}{a_0(e^2_\mathrm{gr}-e^2)^{1/2}},
\end{equation}
in which we have suppressed the divergence by introducing a parameter $e_\mathrm{gr}>1$. The dimensionless parameter
\begin{equation}
\label{eq:epsgr}
\epsilon_\mathrm{gr}\equiv \frac{GM_\bullet}{c^2a_0}\frac{M_\bullet}{N_0m_0},
\end{equation}
measures the relative strengths of apsidal precession due to self-gravity and general relativity. We then replace $\langle\Phi\rangle$ by $\langle\Phi\rangle+\widetilde{H}_\mathrm{gr}$ in equations such as (\ref{eq:free1}). 

Fig.\ \ref{fig:gr} is the analog of Fig.\ \ref{fig:non-sph}, but with the addition of an approximate relativistic Hamiltonian having $e_\mathrm{gr}=1.1$ and a sequence of values of $\epsilon_\mathrm{gr}$. In contrast to systems with no relativistic precession, the rms eccentricity {\em grows} as the system cools when $\epsilon_\mathrm{gr}\goa 0.1$. A phase transition between a disordered, spherically symmetric state and an ordered state with $|\langle\bfe\rangle|\simeq 1$ is present for all values of relativistic precession, but the phase transition changes from first-order for small $\epsilon_\mathrm{gr}$ (discontinuous change in entropy) to continuous as $\epsilon_\mathrm{gr}$ grows. The inverse temperature at the transition approaches a limit $\beta\simeq 18$ as $\epsilon_\mathrm{gr}$ grows arbitrarily large. 

Much of this behavior can be described physically. For example, the growth in $\langle e^2\rangle$ as the system cools arises because the relativistic Hamiltonian (\ref{eq:hgr}) decreases with increasing eccentricity, in contrast to the self-gravitational Hamiltonian (Fig.\ \ref{fig:eee}) which grows with increasing eccentricity. Thus cold systems prefer low eccentricities when relativistic precession is absent, but high eccentricities when it dominates the precession rate. 

\section{Thermodynamic and dynamical stability}

When the phase transition is continuous, the free energy in the disordered equilibrium is an extremum but not a minimum (i.e., a maximum or a saddle point). Thus the disordered state is subject to a linear thermodynamic instability. 

The onset of this instability can be determined analytically. Equation (\ref{eq:free1}) for the free energy can be rewritten as 
\begin{equation}
g(\beta,\bfA)=-\frac{1}{\beta}\log\int\! \rd\bfnu\,\exp\big[-\beta(H_\mathrm{sph}+H_\mathrm{non-sph})\big],
\label{eq:free5}
\end{equation}
where $H_\mathrm{sph}$ includes the Hamiltonian ${H}_\mathrm{gr}$ that describes relativistic precession (if any) and all terms involving the order parameters $A_{00}$, while $H_\mathrm{non-sph}$ contains all terms involving the order parameters $A_{lm}$ with $l>0$. In the spherically symmetric equilibrium  $H_\mathrm{non-sph}=0$. The equilibrium is linearly stable to non-spherical perturbations\footnote{The spherically symmetric equilibria in this paper are always stable to spherical perturbations, i.e., the extremum of the free energy as a function of $A_{00}$ is always a minimum.}  if it is a local minimum of the free energy for all variations of $\{A_{lm}\}$ subject to the constraints $l>0$ and  $A_{l-m}=(-1)^m A_{lm}^\ast$.  This requires 
\begin{equation}
\frac{Gm_0^2N_0\beta}{\pi a_0(l+\tfrac{1}{2})^2}\frac{\int \rd e^2 \exp(-\beta{H}_\mathrm{sph})\int \sin I dI \rd\omega \rd\Omega \, \max\big[(\mbox{Re\,}W_{lm})^2,(\mbox{Im\,}W_{lm})^2\big]}{(1+\delta_{m0})\int \rd e^2\exp(-\beta{H}_\mathrm{sph})} <1
\end{equation}
for all $l>0$. This condition is always satisfied for negative-temperature equilibria ($\beta<0$). The integral involving $(\mbox{Re}\,W_{lm})^2$ or $(\mbox{Im\,}W_{lm})^2$ can be evaluated using equations (\ref{eq:ortho}) and (\ref{eq:wdef}) to give \begin{equation}
\label{eq:stab}
\frac{2\pi Gm_0^2N_0\beta}{a_0(l+\tfrac{1}{2})^3}\frac{\sum_{n=-l}^l Y^2_{ln}(\tfrac{1}{2}\pi,0)\int \rd e^2 \exp(-\beta {H}_\mathrm{sph})U_n^2(e)}{\int \rd e^2\exp(-\beta{H}_\mathrm{sph})} <1,
\end{equation}
which is independent of $m$, as it must be. 

Numerical exploration of equation (\ref{eq:stab}) shows that when relativistic precession is negligible the thermodynamic equilibria are always stable to perturbations with $l=1$, although the left side of (\ref{eq:stab}) approaches unity as $\beta\to\infty$. These results -- stability for $l=1$ but neutral stability when the density of near-radial orbits vanishes -- are consistent with the behavior of black-hole star clusters with arbitrary radial distributions \citep{tre05}.

Fig.\ \ref{fig:gr} shows that when relativistic precession is significant the rms eccentricity grows as $\beta$ increases. If the eccentricities are close to unity we can replace $U_n(e)$ in equation (\ref{eq:stab}) by the limiting form in equation (\ref{eq:unapprox}) and use the relation $\sum_{l=-n}^n Y^2_{ln}(\tfrac{1}{2}\pi,0)=(l+\tfrac{1}{2})/(2\pi)$ to find the stability condition 
\begin{equation}
\beta < \beta_\mathrm{crit}=\frac{\pi^2a_0(2l+1)^2}{32 Gm_0^2N_0}.
\label{eq:beta}
\end{equation}
The most unstable mode has $l=1$. In the dimensionless units defined by equation (\ref{eq:units}) $\beta_\mathrm{crit}=\frac{9}{16}\pi^3=17.44$, consistent with the limiting behavior seen in Fig.\ \ref{fig:gr} for large $\epsilon_\mathrm{gr}$. 

Thermodynamic instabilities in stellar systems are closely related to dynamical instabilities \citep[e.g.,][]{chav06}, and for some classes of stellar system the two instabilities set in at the same point. A linear analysis of the normal modes of the scale-free systems of this paper shows that a neutral mode is present when the left side of (\ref{eq:stab}) equals unity, suggesting that the inequality (\ref{eq:stab}) is also a criterion for dynamical stability. The physical mechanism that drives this instability is related to that of the radial-orbit instability in galaxies \citep{palmer94,tre05,bt08}. Dynamical instabilities can grow on the secular time-scale $\Omega^{-1} M_\bullet/M_\star$, which is faster than the resonant relaxation time-scale by a factor of order $M_\star/m$. However, the evolution of the cluster towards an equilibrium state that is susceptible to the dynamical instability, in which the stars are mostly on high-eccentricity orbits, still occurs only over the resonant relaxation time-scale. 

\section{Discussion}

\label{sec:disc}

In many black-hole star clusters there is a wide range of radii or semimajor axes where the resonant relaxation time is much shorter than both the two-body relaxation time and the cluster age. In these regions, the cluster should be in a thermal equilibrium state in which the number of stars in each semimajor axis interval is conserved and the phase-space distribution function at a given semimajor axis is proportional to $\exp[-\beta m(\langle\Phi\rangle)+H_\mathrm{gr})]$. Here $m$ is the stellar mass, $\langle\Phi\rangle$ is the orbit-averaged potential due to the self-gravity of the stars, $H_\mathrm{gr}$ represents the effects of general relativity, and $\beta$ is the inverse temperature. 
We have explored the thermodynamic equilibria of a self-similar model of such star clusters. The model is completely characterized by two parameters, $\beta$ and the relativistic strength $\epsilon_\mathrm{gr}$ (eq.\ \ref{eq:epsgr}). 

The clusters exhibit a phase transition from a disordered, spherically symmetric, high-temperature equilibrium to an ordered, low-temperature equilibrium in which the stars are on high-eccentricity orbits with similar apoapsis directions. When relativistic effects are negligible, the disordered state is metastable below the critical temperature and the phase transition is first-order. When relativistic effects are important -- roughly, when the apsidal precession due to relativity is faster than the precession due to the self-gravity of the stars -- the disordered state is linearly unstable below the critical temperature and the phase transition is continuous. Although relativistic precession also tends to suppress resonant relaxation (see Fig.\ 4 of \citealt{bof18}), this is a dynamical instability with growth rate comparable to the precession rate, so it persists even when resonant relaxation is slower than two-body relaxation. 

Over time-scales longer than the two-body relaxation time, the semimajor axis distribution will evolve towards the Bahcall--Wolf (1976) distribution, $n(a) \propto a^{-7/4}.$ This process is not captured in our models.

Known results about the stability of star clusters mostly relate to spherical clusters in which relativistic effects are negligible. Collisionless clusters in which the mass is dominated by a central point and the distribution function is monotonic in the angular momentum are dynamically stable to all lopsided ($l=1$) perturbations \citep{tre05}, although in the limiting case of an empty loss cone at zero angular momentum they are only neutrally stable. Such clusters appear to be linearly stable for all spherical wavenumbers $l$ if the distribution function is monotonic in the angular momentum but there is no general proof of this \citep{pps08}.

Our discussion of thermodynamic stability is based on the canonical ensemble, that is, we assume that the stars at each semimajor axis are in thermal equilibrium with a heat bath at temperature $\beta^{-1}$. In practice the bath consists of the stars in the cluster at other semimajor axes, since there is exchange of energy but not stars between semimajor axes. The temperature of the bath is determined by the eccentricity distribution of the stars; in the absence of relativistic effects, smaller rms eccentricity corresponds to lower temperature. For studies of the thermodynamic stability of finite black-hole star clusters it would be more appropriate to use the microcanonical ensemble.

The lifetime of the metastable state is difficult to determine. The simplest estimate \citep[e.g.,][]{chav05} is that the lifetime is of order $t_\mathrm{relax}\exp(\beta N\Delta g)$ where $t_\mathrm{relax}$ is the resonant-relaxation time-scale and $\Delta g>0$ is the
free-energy barrier that the system must cross from the metastable spherical state to the stable ordered state (cf.\ Fig.\ \ref{fig:constrain}). This estimate yields a lifetime that is much longer than the age of the Universe in most cases of interest. However, phenomena such as nucleation or external perturbations can drastically shorten the lifetime.

Our model ignores several important effects. We do not consider consumption or disruption of stars by the central black hole. As we have already argued in \S\ref{sec:sph}, the loss of stars on near-radial orbits to the central black hole cools the system, and the order-disorder phase transition sets in when the temperature is low enough. However, the consumption of stars also erodes the star cluster, and under what conditions the phase transition sets in before the cluster is consumed cannot be decided using the methods in this paper.

We have not considered the effect of a stellar mass distribution.  More massive stars should be concentrated in regions where the gravitational potential is lower. In the disordered state high-mass stars will therefore have lower rms eccentricity than low-mass stars, and in the ordered state the distribution of both eccentricity and apsidal direction will depend on the stellar mass. 

All of our lopsided equilibria are axisymmetric. We have searched for non-axisymmetric lopsided equilibria without success, but a more thorough search is needed before we can rule out the possibility that the generic lopsided maximum-entropy state is non-axisymmetric. 

We have not considered what processes determine the temperature of the cluster. In clusters that are not scale-free, we expect to see a radial temperature gradient that will preferentially cool some regions below the critical temperature.

The evolution of the system in the ordered state is also poorly understood. Presumably the rate of tidal disruption or consumption of stars by the central black hole is higher in this state. These losses might lead to consumption of most or all of the black-hole star cluster in an interval much less than the age of the Universe. 

Our models have zero total angular momentum. We have found a similar phase transition in two-dimensional models of near-Keplerian stellar discs \citep{tt14} but three-dimensional models of rotating discs will ultimately be needed to test whether the ordered thermodynamic equilibrium state can reproduce observations of the lopsided nucleus in M31 \citep{kor99,lauer12}.

This paper complements the results of TTK19, who studied an idealized cluster model in which all the stars had the same semimajor axis, using Markov chain Monte Carlo simulations and direct N-wire integrations of the orbit-averaged equations of motion. These experiments revealed a similar phase transition even though both the numerical tools and the cluster model were very different. Experiments of this kind should be augmented in the future by analyses of black-hole star clusters with more realistic semimajor axis distributions. The most challenging and realistic approach is direct $N$-body simulation  \citep[e.g.,][]{ber18}. 

\section*{Acknowledgements}

This research emerged from discussions with Jihad Touma about instabilities in simulations of black-hole star clusters, and would not have been possible without his insights and encouragement.

\appendix

\section{W- and U- functions}

\label{app:wu}

\noindent
The goal of this appendix is to evaluate the function defined in equation (\ref{eq:w1def}),
\begin{equation}
W_{lm}(e,I,\omega,\Omega)\equiv \langle (a/r)^{1/2}Y_{lm}(\theta,\phi)\rangle = \frac{\big\langle (1+e\cos
   f)^{1/2}Y_{lm}(\theta,\phi)\big\rangle}{(1-e^2)^{1/2}}.
\label{eq:wdef1}
\end{equation}
Here $\langle\cdot\rangle$ denotes a time average over the orbit. Note that
\begin{equation}
\label{eq:cc}
Y_{l-m}(\theta,\phi)=(-1)^m Y^*_{lm}(\theta,\phi), \quad \mbox{and hence}\quad  W_{l-m}(e,I,\omega,\Omega)=(-1)^mW^*_{lm}(e,I,\omega,\Omega).
\end{equation} 
 Using the properties of Kepler orbits it is straightforward to show
that
\begin{align}
\langle X(f) \rangle &=
\frac{(1-e^2)^{3/2}}{2\pi}\int_0^{2\pi}\frac{\rd f}{(1+e\cos
  f)^{2}}X(f), \nonumber \\
\langle X(u) \rangle &=
\frac{1}{2\pi}\int_0^{2\pi}\rd u\,(1-e\cos u)\,X(u), 
\end{align}
where $f$ and $u$ are the true and eccentric anomalies and $X$ is an arbitrary function. We use the representation of a spherical harmonic in orbital elements, 
\begin{equation}
Y_{lm}(\theta,\phi)=\sum_{n=-l}^{\,l}
\ri^{n-m} Y_{ln}(\tfrac{1}{2}\pi,0)  D^{\,l}_{nm}(\Omega,I,\omega)\re^{\ri nf}.
\label{eq:ylm}
\end{equation}
Here 
\begin{equation}
D^{\,l}_{nm}(\Omega,I,\omega)=\re^{\ri n\omega}d_{nm}^{\,l}(I)\re^{\ri m\Omega},
\end{equation}
is the Wigner D-matrix, with 
\begin{equation}
d^{\,l}_{nm}(I)=\sum_s\frac{(-1)^s\sqrt{(l+n)!(l-n)!(l+m)!(l-m)!}}{(l+m-s)!(l-n-s)!s!(s+n-m)!} \left(\cos\tfrac{1}{2}
I\right)^{2l+m-n-2s}\left(\sin\tfrac{1}{2} I\right)^{2s+n-m},
\end{equation}
where the sum is over all integer values of $s$ for which the arguments of the factorials are non-negative. Note that $Y_{ln}(\tfrac{1}{2}\pi,0)=0$ unless $l-n$ is even.

We shall use the orthogonality relation
\begin{equation}
\int_0^{2\pi}\rd\Omega \int_0^\pi \sin I dI \int_0^{2\pi} \rd\omega\, D^{l_1\ast}_{n_1m_1}(\Omega,I,\omega) D^{l_2}_{n_2m_2}(\Omega,I,\omega)=\delta_{m_1m_2}\delta_{n_1n_2}\delta_{l_1l_2}\frac{8\pi^2}{2l_1+1}.
\label{eq:ortho}
\end{equation}

Using these results,
\begin{equation}
W_{lm}(e,I,\omega,\Omega)=\sum_{n=-l}^{\,l}\ri^{n-m}Y_{ln}(\tfrac{1}{2}\pi,0)D^{\,l}_{nm}(\Omega,I,\omega) U_n(e)
\label{eq:wdef}
\end{equation}
where
\begin{equation}
U_n(e)\equiv \frac{1-e^2}{\pi}\int_0^\pi \frac{\rd f\cos nf}{(1+e\cos f)^{3/2}}.
\end{equation}
The limiting behaviours of this function are
\begin{align}
 U_n(e) &\to \frac{(-1)^n \Gamma(|n|+\frac{3}{2})}{2^{|n|-1}\sqrt{\pi}\,\Gamma(|n|+1)}e^{|n|} \quad\mbox{as $e\to
   0$}, \nonumber \\& \to \frac{2^{3/2}(-1)^n}{\pi}\qquad\qquad\qquad \ \mbox{as $e\to 1$}.
   \label{eq:unapprox}
\end{align}
For $n=0$ we have
\begin{equation}
U_0(e)=\frac{2}{\pi}\sqrt{1+e}\,E\bigg(\sqrt{\frac{2e}{1+e}}\,\bigg),
\label{eq:u0}
\end{equation}
where $E$ denotes the complete elliptic integral. Analytic expressions can also be derived for $n>0$, but for repetitive numerical work it is more efficient to evaluate $U_n(e)$ by interpolation in a lookup table. 

\section{The free energy}

\label{app:free}

\noindent
For simplicity, we assume in this Appendix that any deviations of the density or gravitational potential from a power law are periodic in $\log r$ with period $\Gamma$ (eventually $\Gamma$ will disappear from our formulae). Then the density and potential expansion in equations (\ref{eq:rhopot}) can be generalized to a complete set of density-potential basis functions:
\begin{equation}
\rho_{klm}(\bfr)=r^{2\pi \ri k/\Gamma-5/2}\,Y_{lm}(\theta,\phi),
\quad
\Phi_{klm}(\bfr)=-X_{klm}\,r^{2\pi \ri k/\Gamma-1/2}\,Y_{lm}(\theta,\phi)
\label{eq:basis}
\end{equation}
where $k$ is an integer and 
\begin{equation}
X_{klm}\equiv \frac{4\pi G}{(l+\tfrac{1}{2})^2+(2\pi k/\Gamma)^2}.
\end{equation}
The basis functions satisfy the relations 
\begin{equation}
\nabla^2\Phi_{klm}=4\pi G\rho_{klm}, \quad \int_\Gamma \rd\bfr\, \Phi_{klm}^*(\bfr)\rho_{k'l'm'}(\bfr)
=-\Gamma X_{klm}\,\delta_{kk'}\delta_{ll'}\delta_{mm'}.
\label{eq:one}
\end{equation}
Here $\int_\Gamma$ denotes an integral over a spherical annulus with inner and outer radii differing by a factor $\re^{\Gamma}$. The periodicity of the density and potential imply that a star with semimajor axis $a$ and mass $m$ is accompanied by image stars having semimajor axes $\Gamma^na$, masses $\Gamma^{n/2}m$, and the same eccentricities and angular orbital elements, for all integer $n$ (cf.\ eq.\ \ref{eq:ss}). Note that the image stars have different orbital periods proportional to $\Gamma^{3n/2}$ so this model cannot be used to study dynamical evolution, only thermodynamic equilibrium.

The orbit-averaged density and potential of star $j$ and its images can be expanded as 
\begin{equation}
\rho(\bfr,j)=\sum_{klm} a_{klm}(j)\rho_{klm}(\bfr) \quad;\quad  \Phi(\bfr,j)=\sum_{klm} a_{klm}(j)\Phi_{klm}(\bfr).
\end{equation}
The second of relations (\ref{eq:one}) can be used to show that
\begin{equation}
a_{klm}(j)=-\frac{1}{\Gamma X_{klm}}\int_\Gamma \rd\bfr\,
\Phi^*_{klm}(\bfr)\rho(\bfr,j)=-\frac{m_j}{\Gamma X_{klm}}\langle
\Phi^*_{klm}(\bfr)\rangle_j
\label{eq:adef}
\end{equation}
where $m_j$ is the mass of star $j$, and $\langle\cdot\rangle_j$ denotes the time average over its orbit. 

The Hamiltonian associated with the self-gravity of a set of $N$ stars is
\begin{equation}
H=\tfrac{1}{2}\sum_{\genfrac{}{}{0pt}{}{ij=1}{i\not=j}}^N \int_\Gamma \rd\bfr\,\Phi(\bfr,i)\rho(\bfr,j)=-\tfrac{1}{2}\Gamma\sum_{klm}X_{klm}\sum_{\genfrac{}{}{0pt}{}{ij=1}{i\not=j}}^N a_{klm}^*(i)a_{klm}(j)
=-\tfrac{1}{2}\Gamma\sum_{klm} X_{klm} \Big|\sum_{j=1}^N a_{klm}(j)\Big|^2.
\label{eq:ham}
\end{equation}
In the last line we have incorrectly included self-interaction terms such as $|a_{klm}(j)|^2$ but the resulting fractional error is only of order $N^{-1}$ and therefore negligible.  

We now apply the Hubbard--Stratonovich transformation to a subset $S$ of the basis functions (the `active' basis functions). Consider the identity 
\begin{equation}
1=\prod_{klm\in S} \frac{\beta \Gamma X_{klm}}{2\pi}
\int_{-\infty}^\infty   \rd A^R_{klm}\int_{-\infty}^\infty \rd A^I_{klm}\,\exp\Big[-\tfrac{1}{2}\beta \Gamma X_{klm}\Big|\sum_{j=1}^N
  a_{klm}(j)-A_{klm}\Big|^2\Big];
\end{equation}
here $A^R_{klm}=\mbox{Re}(A_{klm})$ and $A^I_{klm}=\mbox{Im}(A_{klm})$. The identity is valid so long as $\beta\Gamma X_{klm}>0$, which is always true since we assume that the temperature is positive. It can be rewritten as
\begin{equation}
\exp(-\beta H)=\exp(-\beta H')\!\!\prod_{klm\in S}  \frac{\beta \Gamma X_{klm}}{2\pi}\int \rd A^R_{klm} \rd A^I_{klm}\,\exp\Big\{\beta \Gamma X_{klm} \mbox{\,Re}\big[A_{klm}\sum_{j=1}^N a^*_{klm}(j)\big]-\tfrac{1}{2}\beta \Gamma X_{klm}|A_{klm}|^2\Big\}.
\end{equation}
where
\begin{equation}
H'\equiv -\tfrac{1}{2}\Gamma \sum_{klm\not\in S}X_{klm}\Big|\sum_{j=1}^Na_{klm}(j)\Big|^2
\end{equation}
is the Hamiltonian associated with the passive basis functions. Our goal is to choose the active basis functions to reproduce most of the actual behavior of the potential. Then $|H'|$ will be small and we can assume that $\exp(-\beta H')\simeq 1$, which we do from now on to keep the expressions compact. 

Apart from constant factors, the partition function is
\begin{align}
\mathfrak{Z}&\propto \int \rd\bfnu_1\cdots \rd\bfnu_N\,\exp(-\beta
  H) \nonumber \\
&\propto \prod_{klm\in S}  \int \rd A^R_{klm} \rd A^I_{klm} \int
  \rd\bfnu_1\cdots \rd\bfnu_N\,\exp\Big\{\beta
  \Gamma X_{klm}\mbox{\,Re}\big[A_{klm}\sum_{j=1}^N a_{klm}^*(\bfnu_j)\big]-\tfrac{1}{2}\beta
  \Gamma X_{klm}|A_{klm}|^2\Big\}.
\end{align}
Here the volume element $\rd\bfnu_j$ is defined in equation (\ref{eq:nudef}). The coordinates $\bfnu_j$ and their integration limits are independent of semimajor axis, so the integral over $\rd\bfnu_1\cdots \rd\bfnu_N$ is a product of $N$ identical integrals. Thus
\begin{equation}
\mathfrak{Z} \propto \prod_{klm\in S} \int  \rd A^R_{klm} \rd A^I_{klm}\, \exp\big[-\beta N g(\beta,\bfA)\big]
\label{eq:zdef}
\end{equation}
where the free energy is 
\begin{equation}
g(\beta,\bfA)=-\frac{1}{\beta}\log \int \rd\bfnu\,\exp\bigg[\beta \Gamma \sum_{{klm}\in S} X_{klm}
\Big(\mbox{\,Re}[A_{klm} a_{klm}^*(\bfnu)]-\frac{|A_{klm}|^2}{2N}\Big)\bigg].
\label{eq:free34}
\end{equation}
In fact this is an upper limit to the free energy because $H'$ is negative-definite, but if the set of active basis functions is well-chosen the limit should be nearly saturated. 

In the thermodynamic limit $N\to\infty$ we rescale $\beta\sim 1/N$ and $A_{klm}\sim N$.  Then $\beta g$ is invariant from (\ref{eq:free34}) and the argument of the exponential in (\ref{eq:zdef}) scales as $N$. Thus the integral in (\ref{eq:zdef}) is dominated by the minimum of the free energy:
\begin{equation}
\log \mathfrak{Z} = -\beta N \min_\bfA g(\beta,\bfA) + \mbox{O}(\log N)
= \beta N g[\beta,\bfA^\mathrm{eq}(\beta)] + \mbox{O}(\log N);
\end{equation}
in words, the equilibrium order parameters $\bfA^\mathrm{eq}(\beta)$ are found at the minima of the free energy. The one-particle partition function is
\begin{equation}
Z(\beta)\equiv \mathfrak{Z}^{1/N}\simeq \exp\big\{-\beta
g[\beta,\bfA^\mathrm{eq}(\beta)]\big\}.
\label{eq:z1def}
\end{equation}

Since we are interested in a scale-free system we now restrict the set of active basis functions to those with $k=0$, and suppress the index $k$. From equations (\ref{eq:basis}) and (\ref{eq:adef})
\begin{equation}
a_{lm}(\bfnu)=\frac{m}{\Gamma}\langle r^{-1/2}Y^*_{lm}(\theta,\phi)\rangle =
\frac{m_0}{\Gamma a_0^{1/2}}W^*_{lm}(e,I,\omega,\Omega)
\end{equation}
where $W_{lm}(e,I,\omega,\Omega)$ is defined by equation (\ref{eq:wdef1}). Therefore the free energy is
\begin{equation}
g(\beta,\bfA)=-\frac{1}{\beta} \log \int \rd\bfnu \,\exp[-\beta
\frakH(e,I,\omega,\Omega;\bfA)]
\label{eq:free7}
\end{equation}
where 
\begin{equation}
\frakH(e,I,\omega,\Omega;\bfA) =\sum_{lm}\frac{4\pi G}{(l+\tfrac{1}{2})^2}
\bigg\{-\frac{m_0}{a_0^{1/2}}\mbox{\,Re}\big[A_{lm}
W_{lm}(e,I,\omega,\Omega)\big]+\frac{\Gamma|A_{lm}|^2}{2N}\bigg\}=\sum_{lm}\frac{4\pi G}{(l+\tfrac{1}{2})^2}
\bigg\{-\frac{m_0}{a_0^{1/2}}\mbox{\,Re}\big[A_{lm}
W_{lm}(e,I,\omega,\Omega)\big]+\frac{|A_{lm}|^2}{2N_0}\bigg\};
\label{eq:hdef}
\end{equation}
in the second equation we have used the first of equations (\ref{eq:ss}) to replace $N$ by $\int_\Gamma \rd N=N_0\int_\Gamma \rd a/a=N_0\Gamma$. 

By differentiating equation (\ref{eq:free7}) with respect to $A^R_{lm}$ and $A^I_{lm}$, we find that the extrema of the free energy occur when the order parameters are
\begin{equation}
A_{lm}^\mathrm{eq}=\frac{N_0m_0}{a_0^{1/2}\int \rd\bfnu\,\exp[-\beta \frakH(e,I,\omega,\Omega;\bfA^\mathrm{eq})]}\int \rd\bfnu\, W^*_{lm}(e,I,\omega,\Omega)\exp[-\beta \frakH(e,I,\omega,\Omega;\bfA^\mathrm{eq})].
\end{equation}
In this equation the term proportional $|A_{lm}|^2/N$ in the Hamiltonian $\frakH$ can be dropped, since it cancels from the numerator and denominator. Therefore the extrema occur at \begin{equation}
A_{lm}^\mathrm{eq}=\frac{N_0m_0}{a_0^{1/2}\int \rd\bfnu\, \exp(-\beta m\langle\Phi\rangle)}\int \rd\bfnu\, W^*_{lm}(e,I,\omega,\Omega)\exp(-\beta m\langle\Phi\rangle), 
\label{eq:aeqdef}
\end{equation}
where $m\langle\Phi\rangle$ is given by equation (\ref{eq:a1}) or (\ref{eq:a111}), with the sum restricted to the active basis functions. This result is the same as equation (\ref{eq:a2}), which was derived by a simpler but less general method. 

It is useful to define the probability distribution function at fixed semimajor axis to be\footnote{This differs from our earlier definition of the number distribution function $F$ (eq.\ \ref{eq:df})  by a factor $\propto a^{-3/2}$.} 
\begin{equation}
f(e,I,\omega,\Omega;\beta)\equiv \frac{1}{Z(\beta)} \exp\{-\beta \frakH[e,I,\omega,\Omega;\bfA^\mathrm{eq}(\beta)]\}.
\end{equation}
Here
\begin{equation}
Z(\beta)=\int \rd\bfnu\, \exp\{-\beta \frakH[e,I,\omega,\Omega;\bfA^\mathrm{eq}(\beta)]\}
\end{equation}
is the partition function defined in equation (\ref{eq:z1def}). The energy of the system at a given semimajor axis may be defined as
\begin{equation}
E(\beta)=\int \rd\bfnu\,f(e,I,\omega,\Omega;\beta)\,\frakH[e,I,\omega,\Omega;\bfA^\mathrm{eq}(\beta)],
\label{eq:energydef}
\end{equation}
and the entropy as 
\begin{equation}
S(\beta)=-\int \rd\bfnu\, f(e,I,\omega,\Omega;\beta) \log f(e,I,\omega,\Omega;\beta).
\label{eq:sdef}
\end{equation}
Recalling that $\p g(\beta,\bfA)/\p\bfA=0$ at $\bfA=\bfA^\mathrm{eq}$, it is straightforward to verify the usual thermodynamic relations
\begin{equation}
Z=\exp(-\beta g), \quad E=-\frac{\rd\log Z}{\rd\beta}=\frac{d}{\rd\beta}(\beta g), \quad S=-\beta^2\frac{d}{\rd\beta}\frac{\log Z}{\beta}=\beta^2\frac{dg}{\rd\beta}, \quad g=E-S/\beta,
\label{eq:thermo}
\end{equation}
where $g(\beta)\equiv g[\beta,\bfA^\mathrm{eq}(\beta)]$. 

%% \bsp
\label{lastpage}
\end{document}